\documentclass[12pt,preprint]{aastex}
\newcommand{\sect}[1]{\S\,\ref{#1}}
\newcommand{\mzams}{\ensuremath{{\cal M}_{\rm ZAMS}}}
\newcommand{\nezw}{\ensuremath{^{22}\mem{Ne}}}

\newcommand{\nadr}{\ensuremath{^{23}\mem{Na}}}
\newcommand{\mgvi}{\ensuremath{^{24}\mem{Mg}}}
\newcommand{\mgfu}{\ensuremath{^{25}\mem{Mg}}}
\newcommand{\mgse}{\ensuremath{^{26}\mem{Mg}}}
\newcommand{\ose}{\ensuremath{^{16}\mem{O}}}
\newcommand{\osi}{\ensuremath{^{17}\mem{O}}}

\newcommand{\nezwa}{\ensuremath{^{20}\mem{Ne}}}

\newcommand{\mem}[1]{\ensuremath{\mathrm{ #1}}}
\newcommand{\msun}{\ensuremath{\, {\cal M}_\odot}}
\newcommand{\be}{\begin{displaymath}}
\newcommand{\ee}{\end{displaymath}}
\newcommand{\bea}{\begin{eqnarray}}
\newcommand{\eea}{\end{eqnarray}}
\newcommand\msol{{\cal M_{\odot}}}

\newcommand\mstar{{\cal M}}

\shortauthors{Denissenkov \& Herwig}
\shorttitle{The Abundance Evolution of Oxygen, Sodium and Magnesium}

\begin{document}

\title{THE ABUNDANCE EVOLUTION OF OXYGEN, SODIUM AND MAGNESIUM IN
       EXTREMELY METAL-POOR INTERMEDIATE MASS STARS: IMPLICATIONS
       FOR THE SELF-POLLUTION SCENARIO IN GLOBULAR CLUSTERS}

\author{Pavel A. Denissenkov \& Falk Herwig} 
\affil{Department of Physics \& Astronomy, University of Victoria, 
       P.O.~Box 3055, Victoria, B.C., V8W~3P6, Canada}
\email{dpa@uvastro.phys.uvic.ca, fherwig@uvastro.phys.uvic.ca}
 
\begin{abstract}
We present full stellar evolution and parametric models of
the surface abundance evolution of \ose, \nezw, \nadr, and the
magnesium isotopes in an extremely metal-poor intermediate mass star
($\mzams = 5\msun$, $Z=0.0001$). \ose\ and \nezw\ are injected into the
envelope by the third dredge-up following thermal pulses on the
asymptotic giant branch. These species and the initially present
\mgvi\ are depleted by hot bottom burning (HBB) during the interpulse
phase. As a result, \nadr, \mgfu\ and \mgse\ are enhanced. If the HBB
temperatures are sufficiently high for this process to deplete oxygen
efficiently, \nadr\ is first produced and then depleted during the
interpulse phase. Although the 
simultaneous depletion of \ose\ and enhancement of \nadr\ is
possible, the required fine tuning of the dredge-up and HBB
casts some doubt on the robustness of this process as the origin of 
the O--Na anti-correlation observed in globular cluster stars. However, a very
robust prediction of our models are low \mgvi/\mgfu\ and  \mgvi/\mgse\
ratios whenever significant \ose\ depletion can be
achieved. This seems to be in stark contrast with recent observations
of the magnesium isotopic ratios in the globular cluster NGC~6752.
\end{abstract} 

 
\section{Introduction}
\label{sec:intro}
In globular clusters (GCs) spectroscopic observations have revealed large ($\sim 1$\,dex)
star-to-star abundance variations of C, N, O, Na, Mg and Al
\citep[e.g.][]{rc02}. The anti-correlations of C--N, O--Na and Mg--Al
point to simultaneous operation of
the CNO-, NeNa-, and MgAl-cycles. 
In the so-called \emph{evolutionary scenario} it is assumed that
these abundance variations are produced in the vicinity of 
the hydrogen-burning shell in red giant branch (RGB) stars and that
some extra-mixing transports them to the convective envelope
(\citealt{dd90,lhs93,dw01}).

The evolutionary scenario has been challenged by the  recent 
discovery of C--N, O--Na and even Mg--Al anti-correlations in the main sequence (MS),
MS turn-off and subgiant stars in the GCs 47~Tucanae and NGC~6752
\citep{hea03,grea01,gea02}. A likely origin of these abundance
anomalies is pollution by material processed via
H-burning in more evolved stars.  In this \emph{primordial scenario}
intermediate-mass stars 
(IMS, $\mstar\approx 3$--$8\msol$) in their asymptotic giant branch (AGB)
evolution phase 
have been proposed as contaminators (\citealt{vea01,vea02}).

Thermally pulsing (TP-) AGB stars process material in a He- and a
H-burning shell. The H-shell dominates the energy production most of the
time. However, recurrent thermonuclear He-shell flashes drive  a
temporary  ($\sim$\,10\,yr) pulse-driven convective zone (PDCZ) that
encompasses the entire region between the He- and the H-burning shell
(the intershell). Immediately after the end of the TP the base
of the convective envelope begins to move inwards in mass and
eventually material from below the H-shell is dredged-up to the
envelope (third dredge-up, TDU). During the interpulse period in
massive AGB stars, hot-bottom burning (HBB)  further modifies  
the envelope chemical composition \citep{boothroyd:93}. 

In the evolutionary scenario Al can be synthesized on the RGB
if some low-mass stars in GCs were
initially enriched with $^{25}$Mg, possibly from massive AGB stars
\citep{dea98,dw01}. In this {\it combined scenario}
star-to-star abundance variations of C, N, O, Na, Mg and Al
in GCs may have multiple origins: {\it (i)} 
on the MS they may be
due to pollution in the past either by the massive AGB stars
or by somewhat more massive ($0.9\la\mstar/\msol\la 2$) RGB stars than 
the present-day MS turn-off stars which had undergone deep extra-mixing, 
and {\it (ii)}  on the RGB they may
be partly tracers of the same pollution that occured 
on the MS and partly (in the most rapidly rotating stars) 
they may be due to deep extra-mixing \citep{dw01,dv03}.

In the primordial scenario massive AGB stars are thought to be responsible for very
low O abundances ([O/Fe]\,$\la -0.5$, down from the assumed initial
value of +0.4) in MS stars in GCs by \emph{pollution with O-depleted material}.
Indeed,  \citet{vea01,vea02} have reported that, in metal-poor
massive AGB stars, HBB may be  capable of producing
the required O depletion. However, nucleosynthesis beyond 
the CNO-cycle has not been investigated yet. In this {\it Letter} we
will take into account all of the cycles of 
nuclear reactions participating in the H-burning as
well as the effect of the TDU. In \sect{sec:fullSE} we
present the abundances from full stellar evolution models. In
\sect{sec:param} an equivalent parametric AGB model is described.
In \sect{sec:concl} we discuss the results of calculations with the parametric models
and make our final conclusions.

\section{Full Stellar Evolution Models}
\label{sec:fullSE}
Our 1D stellar evolution code \citep{herwig:99a} includes
updated opacities \citep{iglesias:96,alexander:94} and a nuclear
network with all relevant reactions with the rates for the NeNa-cycle reactions taken 
from \citet{eech95}. Simultaneous, fully implicit, iterative solution of 
the nuclear network and time-dependent mixing equations 
for each isotope \citep{herwig:00a}, and hydrodynamic
overshooting with a geometric, exponential decay parameter $f$ can be
included. We have improved the adaptive time-step and
grid-allocation algorithm for the extremely metal-poor models in order
to ensure that the TDU properties are not affected by
numerical resolution issues. The mixing-length parameter is
$\alpha_\mem{MLT}=1.7$ from calibrating a solar model, 
and $X(^4\mathrm{He})_\mathrm{init}=0.23$. 

We choose a model of initially $5\msun$ with a metallicity of
$Z=0.0001$ as a representative example for IMS
that might have polluted GC stars of the lowest
metallicity ([Fe/H]\,$\approx\log(Z/Z_\odot)$\,=\,$-2.3$). 
We evolve the initial pre-MS model through all evolutionary
phases, and details will be presented in a forthcoming paper. Up to
the first thermal pulse on the AGB, exponential 
overshooting with $f=0.016$ has been considered at all convective 
boundaries. The first envelope
abundance alteration occurs as a result of the second 
dredge-up. Material processed mainly by
H-shell burning is brought to the surface. \nadr\ is enhanced by
$0.75\,\mem{dex}$ from the conversion of the initial abundance of \nezw\
and some \nezwa. \ose\  is depleted by $0.1\,\mem{dex}$ due to the action of 
the ON cycle. Magnesium isotopes are changed by less than $0.05$,
$0.1$ and $0.01\,\mem{dex}$ for mass numbers $24$, $25$ and $26$ in the
second dredge-up. 

During the TP-AGB phase the interplay of the TDU
and HBB alters the envelope abundances. 
TDU brings \nezw\ and a rather uncertain amount of
\ose\ to the surface. \mgfu\ and \mgse\ are produced by two processes:
{\it (i)} by HBB in the envelope (at the expense of $^{24}$Mg), and {\it (ii)}
by $\alpha$ captures on \nezw\ in the PDCZ. The second process is important
when the temperature in the PDCZ exceeds $\sim$\,$3.5\times 10^8$\,K.
The HBB also destroys \ose\ (and produces \osi),
while \nadr\ is produced by proton captures on the dredged-up
\nezw. At higher temperatures, \nadr\ can be
destroyed again. The  overall budget of the O, Na and Mg
isotopes depends on the interplay of dredge-up and
HBB. In addition, mass loss plays an important role. It
limits the TP-AGB evolution time with high HBB temperature and efficient
third dredge-up.

All of these effects can be observed in  the surface abundance evolution of
two TP-AGB model sequences that we evolved from the same early-AGB
model (Fig.~\ref{fig:f1}). In the first sequence (a) we 
have isolated the effect of HBB by assuming no overshooting
and a rather low time
resolution (typically less than 1000 models per TP cycle). As a result,
this sequence shows no TDU (apart from one \emph{outlier}
seen in the run of \ose\ at $t=60000\,\mem{yr}$). We have also assumed
mass loss according to \citet{bloecker:95a} with $\eta_\mem{B}=0.1$,
and the model experiences 43 TPs until the envelope mass is lost.
This sequence shows a large \ose\ depletion, but due to the lack of
\nezw\ dredge-up, \nadr\ is depleted as well. 

In panel (b) a model is shown in which we have considered only negligible 
overshooting at the bottom of the He-flash convection zone ($f=0.002$;
the intershell abundances are not affected significantly by this very
small PDCZ overshoot). At the bottom of the
envelope convection we have assumed $f=0.016$, as calibrated at the
MS core convection boundary.  Here we have used the high
spatial and  time resolution required to model the TDU. Mass loss has 
been turned off for this case in order to explore the largest possible nuclear
enrichment by HBB and dredge-up. 
After a few TP the TDU is very efficient, and
it reaches into $^{16}$O-rich layers below the zones previously covered by
the PDCZ. Our models with this deep dredge-up show systematically
higher HBB temperatures than the model without
dredge-up. The unusually deep dredge-up is likely related to the
effect of sustained, fierce H-burning in the overshoot layer. We will
discuss the details of this effect, which is unique to very metal
poor IMS, in a forthcoming paper. For the present study it is sufficient
to note that the TDU in these stars could be much more
efficient than previous models without overshooting indicated
\citep[e.g.][]{vea02}, and that oxygen could be enhanced by this
process. This undermines the ability of HBB
to deplete oxygen in the envelope. However, with efficient
TDU, \nezw\ is dredged-up and \nadr\ is, in fact, further increased. The behaviour of
the magnesium isotopes is qualitatively the same as in the
model without dredge-up, indicating that these species are mainly
affected by the HBB.

According to our high-resolution, full stellar evolution models with
efficient TDU, massive AGB stars {\it cannot} show simultaneously
$^{16}$O depletion and Na enhancement. Using parametric models we
will now explore intermediate cases with \nezw\ dredge-up but limited
\ose\ dredge-up, and we will consider in more detail the evolution of
magnesium isotopes.

\section{Parametric Models}
\label{sec:param}
We have developed a parametric nucleosynthesis and mixing
code  for massive AGB stars. All processes that are relevant for the abundance
evolution of O, Na, and the Mg isotopes are considered. In particular,
we include the effects of HBB and dredge-up, as well as mixing and burning
in the PDCZ. Most reaction rates are taken from the NACRE  \citep{aea99} 
compilation \citep[see][for details]{dt03}. 
For the initial conditions we assume scaled solar
abundances except for an enhancement of the $\alpha$-elements ($^{16}$O, $^{20}$Ne,
$^{24}$Mg, etc.) by $+0.4$\,dex, and a depletion of
Na and Al by the same factor, in accordance with the chemical composition of 
the halo dwarfs. As in the full stellar models, the metallicity is
$Z=0.0001$.  In the following description,
we use the subscripts ``PDCZ'', ``HBS'' and ``HBB'' to indicate the
temperatures and densities at the base of the PDCZ, in the
H-burning shell and at the bottom of the convective envelope, respectively.  

A TP sequence starts with a calculation of the mixing and
nucleosynthesis in the PDCZ. 
The pre-TP composition
of the intershell zone is a mixture consisting of a fraction $\gamma$ of material from
the preceding PDCZ and a fraction $(1-\gamma)$ containing H-shell ashes.
For the first TP we take $\gamma = 0$.
The mass $\Delta{\cal M}_{\rm PDCZ}$ of the PDCZ is divided into 25 zones with the temperature and density 
decreasing linearly from $T_{\rm PDCZ}$ and $\rho_{\rm PDCZ}$ at its bottom to
$T_{\rm HBS}$ and $\rho_{\rm HBS}$ at its top. For the mixing in the PDCZ,
which is treated as a diffusion process,
we assume a constant coefficient $D_{\rm mix}=10^{15}$\,cm$^2$\,s$^{-1}$. For the temperature
$T_{\rm PDCZ}$  we choose the maximum value observed in the full
stellar models. For that reason we are really modeling only the
high-temperature phase of the PDCZ.
The He-burning is stopped every time when the mass fraction abundance of
$^{12}$C in the PDCZ has reached the value 0.23, as proposed
by \citet{rv81}.
After that a fraction $\lambda$ of the total mass $\Delta{\cal M}_{\rm PDCZ}$ of
material with the final abundances from the PDCZ is added to the
envelope to simulate the effect of the TDU.
Then, we follow the changes of the envelope composition due to the HBB
during the interpulse period 
$\Delta t_{\rm ip}$. For this, we process the envelope abundance
distribution after the TDU in one zone at constant $T_{\rm
HBB}$ and $\rho_{\rm HBB}\delta$ 
(a dimensionless factor $\delta$ accounts for the fact that,
after averaging over the mass of the convective envelope,
coefficients in the nuclear kinetics network can be written in a form
$\langle\sigma v\rangle(T_{\rm HBB})N_{\rm A}\rho_{\rm HBB}\delta$,
where $\langle\sigma v\rangle(T)N_{\rm A}$ is a reaction rate at
a temperature $T$ and $\delta = 
(1-a)^{-1}\int^1_a(\rho/\rho_{\rm HBB})[\langle\sigma v\rangle(T)/
\langle\sigma v\rangle(T_{\rm HBB})]dx$ with $x=M_r/M$ and $a=M_{\rm HBB}/M$;
the full stellar evolution models give $\delta\approx 10^{-5}$, which means that
HBB takes place in a narrow zone adjacent to the base of the convective envelope).
In addition to the HBB computation, we calculate the  
abundance distribution of the H-burning ashes from the new post-HBB envelope
composition with another one-zone model at H-shell temperature and density.
This whole sequence is repeated for eight TP cycles.

We use the following structure parameters from the full models: the
CO-core mass $\mstar_{\rm c}=0.96\msol$,  $\Delta\,{\cal M}_{\rm PDCZ}=0.0025\msun$,
$T_{\rm PDCZ}=3.2\times 10^8$\,K, $\rho_{\rm PDCZ}=5\times
10^3$\,g\,cm$^{-3}$,  
$T_{\rm HBS}=1.2\times 10^8$\,K, $\rho_{\rm HBS}=40$\,g\,cm$^{-3}$, 
$T_{\rm HBB}=10^8$\,K, $\rho_{\rm HBB}\delta =10^{-5}$\,g\,cm$^{-3}$, and
$\Delta t_{\rm ip} = 10^4$\,yr. Our test calculations have shown that the
envelope abundances depend weakly on the parameter $\gamma$ for a wide range of $0.2\leq
\gamma\leq 0.6$. Therefore, its value was kept constant at $\gamma =0.6$, as given by
the full stellar evolution calculations.

The efficiency of nuclear processing of the envelope material in the HBB is determined by 3 parameters:
$\Delta t_{\rm ip}$, $\rho_{\rm HBB}\delta$ and $T_{\rm HBB}$. However, while the efficiency
is linearly proportional to the first two of them, it depends on a
high power of 
$T_{\rm HBB}$. Therefore, in our model only the third parameter
characterizes the efficiency of HBB.
The envelope abundances are also strongly affected by
the efficiency of the TDU (parameter $\lambda$).
In the full stellar evolution models $\lambda$
depends critically on the efficiency of convection induced
extra-mixing, while $T_{\rm HBB}$ depends on the
efficiency of convective energy transport in the envelope. In our
parametric model $\lambda$ and $T_{\rm HBB}$  are free parameters.

\section{Results from the parametric model and general discussion}
\label{sec:concl}
\begin{figure}
\plotone{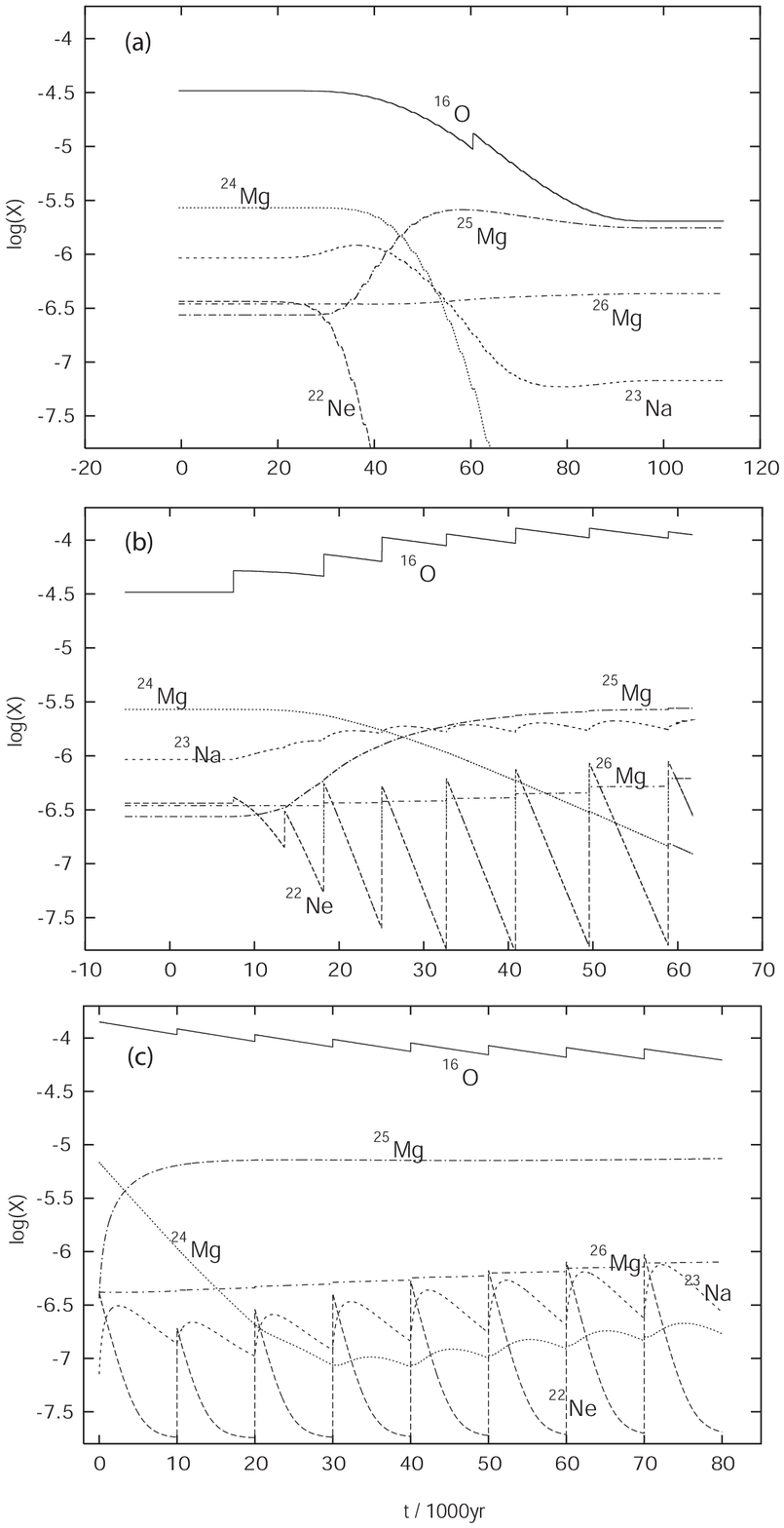}
\caption{Envelope abundance evolution for two full TP-AGB stellar
         model sequences with ${\cal M}_{\rm ZAMS}=5\msun$, $Z=0.0001$ 
         (panels a and b). The sequences are distinguished by differing 
         assumptions concerning the mixing and numerical time resolution,
         as discussed in the text. Panel (c) shows envelope abundance evolution 
         for our parametric AGB model with $T_{\rm HBB}=10^8$\,K and $\lambda =1$. 
         In all panels $t=0$ corresponds to the first TP.} 
\label{fig:f1}
\end{figure}

\begin{figure}
\plotone{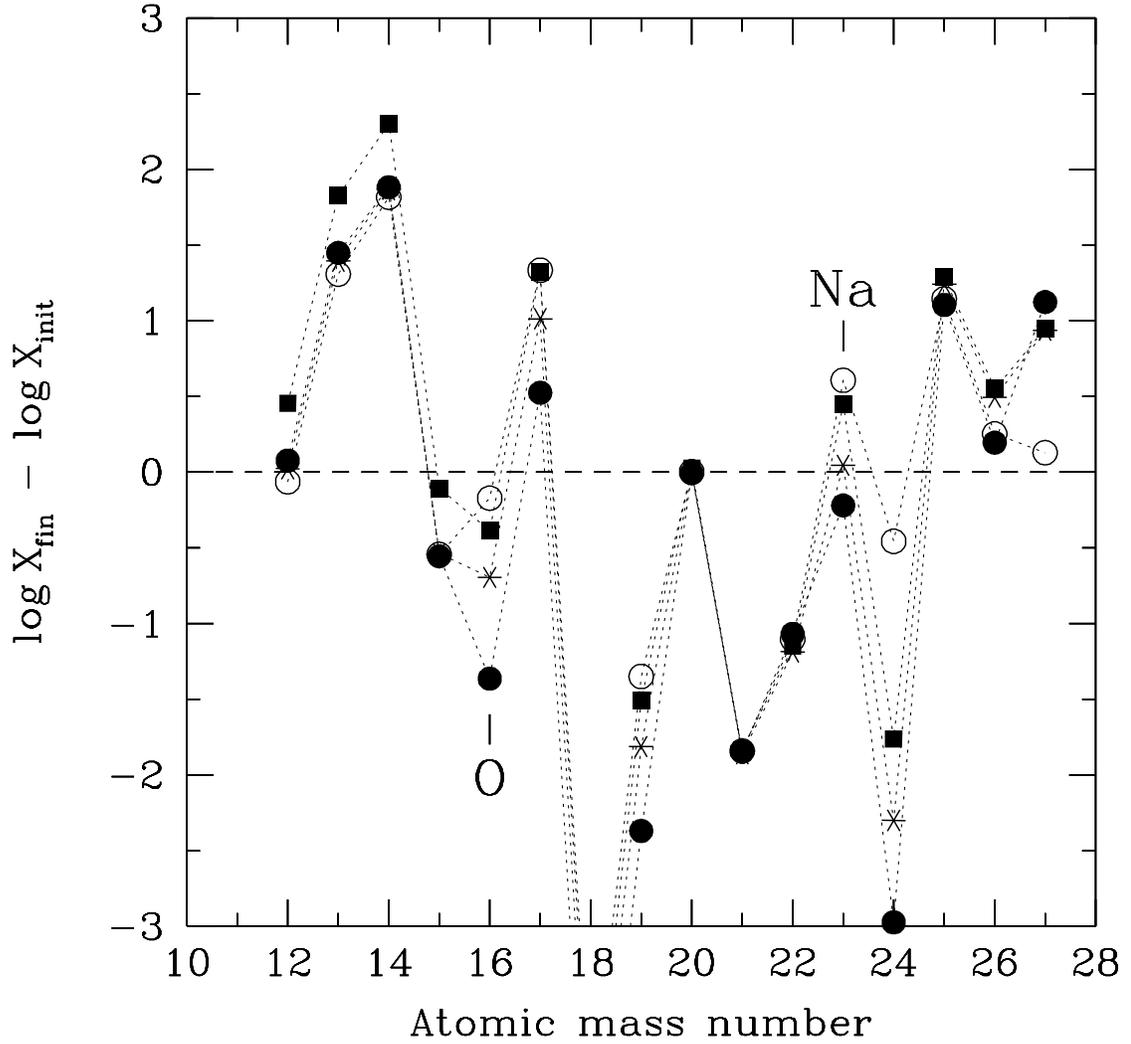}
\caption{Final envelope abundances with respect to the initial ones
         after 8 thermal pulses calculated with the parametric AGB model:  
         $(\lambda,\,T_{\rm HBB})$\,=\,(1,\,$10^8$\,K) ({\it filled squares}),
         (0.3,\,$9\times 10^7$\,K) ({\it open circles}),
         (0.3,\,$10^8$\,K) ({\it asterisks}), and (0.3,\,$1.1\times 10^8$\,K) 
         ({\it filled circles}).
         }
\label{fig:f2}
\end{figure}

Parametric calculations were carried out for $\lambda=0.3$ and $1$, and
$T_{\rm HBB}=0.9 \dots 1.1 \times 10^{8}\mem{K}$. The time evolution
of the envelope abundances (Fig.~\ref{fig:f1}c) resembles the main
features seen in the results of the full stellar evolution calculations:
the dredge-up of $^{22}$Ne, production and destruction of Na via
$^{23}$Na(p,$\alpha)^{20}$Ne and 
$^{23}$Na(p,$\gamma)^{24}$Mg, depletion of \mgvi\ and production of
\mgfu\ and \mgse. \ose\ is depleted just by HBB.
In our standard parametric model we assume that 2\% of the $^{16}$O mass fraction is
dredged-up from the CO--core during every TP, 
in agreement with the average \emph{minimum} predicted by the full
stellar models. 
In the parametric model the $^{16}$O abundance becomes significantly
depleted with  $X_{\rm fin}(^{16}{\rm O})/X_{\rm init}(^{16}{\rm O})\la 0.2$ for
$T_{\rm HBB}> 10^8\mem{K}$ and $\lambda = 0.3$ (Fig.~\ref{fig:f2}).
Thus, if no or very little $^{16}$O is brought to the surface,
then we confirm the findings of \citet{vea01,vea02} that $^{16}$O is efficiently 
destroyed by the HBB in the metal-poor massive AGB stars. However, our
parametric models confirm the result of the full models that high
HBB temperatures do not favour Na production which would be required
to explain the O--Na anti-correlation in GC stars in the primordial
scenario.
Indeed, at $T=10^8$\,K, the sum of the rates of the reactions $^{23}$Na(p,$\alpha)^{20}$Ne
and $^{23}$Na(p,$\gamma)^{24}$Mg
is $\sim$\,6.8 times as large as that of $^{16}$O(p,$\gamma)^{17}$F, the lower and upper
limits of this ratio being 2.1 and 45, respectively
(\citealt{aea99}). Without the \nezw\ dredge-up source which
replenishes Na (as in our first full stellar evolution model), the O
depletion is accompanied by Na destruction at $T=10^8\mem{K}$.

But even with the dredge-up of $^{22}$Ne, the final Na abundance
in the envelope is found to be very sensitive to small variations of $T_{\rm HBB}$:
increasing $T_{\rm HBB}$ from $0.9\times 10^8$\,K to $1.1\times 10^8$\,K turns the Na
production into the Na destruction (Fig.~\ref{fig:f2}). Consequently,
for high HBB temperatures, both  O and Na may be depleted
simultaneously, resulting in O--Na correlation instead of
the O--Na anti-correlation. If dredge-up of \ose\ is as efficient as
predicted by the high-resolution full stellar models with hydrodynamic
overshooting, the temperatures for efficient $^{16}$O depletion would
definitively be too high for Na production.
Furthermore, \citet{iea01} have recommended to use a rate of the reaction
$^{22}$Ne(p,$\gamma)^{23}$Na which is by a factor of $\sim$\,$10^{2.7}$ smaller than
the value given by NACRE at $T=10^8$\,K. This makes Na production in
massive AGB stars even more problematic. On the other hand, a rate of the reaction
$^{22}$Ne(p,$\gamma)^{23}$Na from \citet{eech95}, used in \sect{sec:fullSE},
is by a factor of $\sim$\,$10^{3.9}$ larger than
the NACRE value, which favours Na production.
In summary, we cannot entirely rule out that contamination by massive AGB stars
causes the O--Na anti-correlation in the primordial
scenario. However, this seems to be very unlikely because
it requires a fine tuning of the AGB model parameters.
This is not supported by the latest rate of the reaction
$^{22}$Ne(p,$\gamma)^{23}$Na either.

A very robust prediction of the primordial scenario can be made with
respect to the magnesium isotopes. For temperatures in the vicinity of $T_{\rm
HBB}$\,$\approx$\,$10^8$\,K, that allow \ose\ depletion, the $^{24}$Mg
abundance is depleted in any case even more, producing \mgfu\ and
\mgse\ in turn. This result is consistently found in all of our models for
a wide range of parameters.  The evolution of \mgvi\ in relation to
\ose\ is a robust result because: {\it (i)} at $T$\,=\,$10^8$\,K, the ratio of
the $^{24}$Mg(p,$\gamma)^{25}$Al rate to the
$^{16}$O(p,$\gamma)^{17}$F rate is $\sim 6.4$, the lower and upper
limits being 3.9 and 13 (\citealt{aea99}), and {\it (ii)}
there is no source of $^{24}$Mg in AGB stars (production of $^{24}$Mg
in the reaction $^{23}$Na(p,$\gamma)^{24}$Mg in the HBB is unimportant). 

Therefore, in the primordial scenario  with massive AGB stars
as the contaminators, the MS turn-off stars with  [O/Fe]\,$\la -0.5$ {\it
must} have $^{24}$Mg depleted and $^{25}$Mg enhanced 
by more than one order of magnitude. Our assumed initial chemical
composition has [O/Fe]\,=\,0.4, [$^{24}$Mg/Fe]\,=\,0.4,
[$^{25}$Mg/Fe]\,=\,0.0 and [$^{26}$Mg/Fe]\,=\,0.0, 
i.e. the Mg isotopic ratios $^{24}$Mg\,:\,$^{25}$Mg\,:\,$^{26}$Mg = 90.5\,:\,4.5\,:\,5.0.
According to those models in Fig.~\ref{fig:f2} with significant
O-depletion, material released by massive AGB stars may have $\log(X_{\rm
fin}/X_{\rm init})\approx -1.0,\ -1.5,\ 1.2$ and 0.5 for 
O, $^{24}$Mg, $^{25}$Mg and $^{26}$Mg (in Fig.~\ref{fig:f2} the label $^{26}$Mg represents the sum of
$^{26}$Mg and $^{26}$Al).  In low-mass RGB stars in the primordial scenario,
$\sim$\,90\% of the mass of the convective  
envelope has to consist of this material.
Accordingly, on the upper RGB, such a star would have 
[O/Fe]\,=\,$-0.32$, [$^{24}$Mg/Fe]\,=\,$-0.50$, [$^{25}$Mg/Fe]\,=\,1.2, [$^{26}$Mg/Fe]\,=\,0.5, and
the Mg isotopic ratios $^{24}$Mg\,:\,$^{25}$Mg\,:\,$^{26}$Mg = 13\,:\,71\,:\,16.

These predictions seem to be in conflict 
with the results of Mg isotopic composition analysis of RGB stars in the GC
NGC~6752 reported by \citet{yea03}. In the least polluted stars these authors infer 
$^{24}$Mg\,:\,$^{25}$Mg\,:\,$^{26}$Mg\,$\approx$\,80\,:\,10\,:\,10 and [O/Fe]\,$\approx$\,0.6
while in the most contaminated stars they find
$^{24}$Mg\,:\,$^{25}$Mg\,:\,$^{26}$Mg\,$\approx$\,60\,:\,10\,:\,30 and [O/Fe]\,$\approx -0.1$.
Thus, despite having O depleted by a factor of $\sim$\,5, the NGC~6752 red giants still exhibit 
the $^{24}$Mg dominated isotopic ratios. Moreover, the second most abundant isotope is
$^{26}$Mg instead of $^{25}$Mg. We see only two possibilities to remove
this disagreement within the primordial scenario:
either {\it (i)} the ratio of the reaction rates of $^{24}$Mg(p,$\gamma)^{25}$Al 
and $^{16}$O(p,$\gamma)^{17}$F at $T\approx 10^8$\,K is much
less than the value given by \citet{aea99}, and at the same time
the reaction $^{25}$Mg(p,$\gamma)^{26}$Al is faster; or {\it (ii)} the HBB temperature
in massive metal-poor AGB stars is somewhat lower than $10^8$\,K, in which case
the $^{24}$Mg destruction would be suppressed (see the open circles in Fig.~\ref{fig:f2}).
However, in the second case, $^{16}$O destruction would be suppressed as well, and $^{25}$Mg
(and not $^{26}$Mg) could still be produced in a large amount. Therefore, we would need deep
extra-mixing on the RGB to deplete O and to produce $^{26}$Mg and probably Al (either in the form of
$^{27}$Al or $^{26}$Al$^{\rm g}$) at the expense of $^{25}$Mg. This is exactly the combined scenario
proposed by \citet{dea98}, with the minor modification that deep extra-mixing in the RGB stars
slightly more massive than the present-day MS turn-off stars in GCs might have contributed to
the star-to-star abundance variations as well.

\acknowledgements
We are grateful to the referee Dr. Santi Cassisi for several
thoughtful comments and suggestions that have served to improve this paper.
We appreciate the support from D. A. VandenBerg through his Operating Grant from the Natural Sciences and
Engineering Research Council of Canada.

\clearpage

%-----------------------------------------------------------------
%-----------------------------------------------------------------

\end{document}